\documentclass[iop,apj,tighten]{emulateapj}
\usepackage[breaklinks,colorlinks,citecolor=blue,linkcolor=magenta]{hyperref}
\usepackage{booktabs}
\usepackage{bm}
\usepackage{color}
\shorttitle{Evolution of compact groups}
\shortauthors{Farhang et~al.}

\begin{document}
\title{Evolution of compact and fossil groups of galaxies from semi-analytical models of galaxy formation}
\author{Amin Farhang$^{1}$, Habib G. Khosroshahi$^{1,2}$, Gary A. Mamon$^{2}$, Ali. A. Dariush$^{3}$ and Mojtaba Raouf$^{1}$}
\affil{School of Astronomy, Institute for Research in Fundamental Sciences (IPM), P. O. Box 19395-5746, Tehran, Iran$^{1}$\\
Institut d'Astrophysique de Paris (UMR 7095: CNRS \& UPMC, Sorbonne-Universit\'es), 98 bis Bd Arago, F-75014 Paris, France$^{2}$\\
Institute of Astronomy, University of Cambridge, Madingley Road, Cambridge CB3 0HA, UK$^{3}$\\
{\tt a.farhang@ipm.ir}}

%%%%%%%%%%%%%%%%%%%%%%%%%%%%%%%%%%%%%%%%%%%%%%%%%%%%%
%%%%%%%%%%%%%%%%%%%%%%%%%%%%%%%%%%%%%%%%%%%%%%%%%%%%%
\begin{abstract}
We compare the mean mass assembly histories of compact and fossil galaxy groups in the Millennium dark matter simulation and an associated semi-analytic galaxy formation model. Tracing the halo mass of compact groups (CGs) from $z=0$ to $z=1$ shows that, on average, 55 per~cent of the halo mass in compact groups is assembled since $z\sim 1$, compared to 40 per~cent of the halo mass in fossil groups (FGs) in the same time interval, indicating that compared to FGs, CGs are relatively younger galaxy systems. At $z=0$, for a given halo mass, fossil groups tend to have a larger concentration than compact groups. Investigating the evolution of CG's parameters show that they become more compact with time. CGs at $z=0.5$ see their magnitude gaps increase exponentially, but it takes $\sim$ 10 Gyr for them to reach a magnitude gap of 2 magnitudes. The slow growth of the magnitude gap leads to only a minority ($\sim 41$ per cent) of CGs selected at $z=0.5$ turning into a FG by $z=0$. Also, while three-quarters of FGs go through a compact phase, most fail to meet the CG isolation criterion, leaving only $\sim$ 30 per~cent of FGs fully satisfying the CG selection criteria. Therefore, there is no strong link of CGs turning into FGs or FGs originating from CGs. The relation between CGs and FGs is thus more complex, and in most cases, FGs and CGs follow different evolutionary tracks.

\end{abstract}

%%%%%%%%%%%%%%%%%%%%%%%%%%%%%%%%%%%%%%%%%%%%%%%%%%%%%
%%%%%%%%%%%%%%%%%%%%%%%%%%%%%%%%%%%%%%%%%%%%%%%%%%%%%
\section{INTRODUCTION}
Compact groups of galaxies (CGs) are small galaxy systems in which at least four luminous galaxies are found to be in a compact configuration with a typical inter-galactic separation of the order of scale of the constituent galaxies. Although the first CG was found by \cite{Stephan1877}, however, it was only after the first large survey, i.e. the Palomar Observatory Sky Survey (POSS), that the number of CGs increased substantially. In the mean time, CGs were also catalogd in the Atlas of Interacting Galaxies \citep{VorontsovVelyaminov59,VorontsovVelyaminov77} and the Atlas of Peculiar Galaxies \citep{Arp66_book}. These catalogs contain information on galaxies or galaxy groups selected on the basis of observed signatures of interaction or peculiar appearance. Later studies by \cite{Hickson+77} and \cite{Heiligman&Turner80} identified CGs based on specific, quantitative criteria from the analysis of their morphological features in the POSS photographic imaging plates. These efforts led to publication of the first catalogs of CGs, e.g. the Rose catalog \citep{Rose77} and the Hickson CG catalog \citep[HCG;][]{Hickson82}. More catalogs of CGs were then compiled by applying Hickson's criteria on large-sky surveys such as the Sloan Digital Sky Survey (SDSS, \citealp{Lee+04, McConnachie+09}) and the Two Micron All-Sky Survey (2MASS, \citealp{DiazGimenez+12}).

Studies of CGs indicate that these groups have star formation rates, colors and morphological types that put them somewhere between binary and isolated galaxies \citep{Mamon86,Moles+94,Tovmassian+06}. Based on observations, the median projected galaxy separation in Hickson CGs is approximately $\approx 39\, h^{-1}\,\rm kpc$ with a line-of-sight velocity dispersion of $\approx 200\,\rm km\, s^{-1}$ \citep{Hickson+92}. In such an environment, the dynamical time scale is very short which in turn makes it a commonplace for galaxy-galaxy mergers and interactions. Thus galaxies in CGs tend to be different from the field population, so that the fraction of early-type galaxies in CGs, in magnitude-limited surveys, is significantly higher than in the field. For instance \cite{Hickson&Rood88} found that $\sim$51 per~cent of galaxies in their sample of CGs are early-types, compared to $\sim$20 per~cent in the field \citep{Gisler80}. Also, \cite{Carnevali+81}, \cite{Barnes85} and \cite{Mamon87} studied short dynamical and galaxy merging times within dense groups. As early-type galaxies can be formed via the mergers of late-type systems \citep[e.g.][]{Barnes90}, these observations can be explained by frequent interactions and mergers among galaxies in the environment of CGs \citep{Mamon86, Mamon87}.

\cite{TorresFlores+13} found that the Tully-Fisher relation in galaxies belonging to CGs is similar to those found for field galaxies. \cite{Coenda+12} found that the galaxies in CGs are more concentrated, have higher surface brightness and are smaller in size than galaxies in the field and loose groups. Similarly, \cite{Martinez+13} found that brightest group galaxies (BGGs) in CGs are more concentrated and have a larger surface brightness than their counterparts in both high and low-mass loose groups. \cite{Sohn+13} studied the activity in galactic nuclei in CGs and found a strong (respectively weak) environmental dependency of AGN fraction for early-type (late-type) galaxies in CGs. \citeauthor{Coenda+12} also found that, while the luminosity function of galaxies in CGs has a characteristic magnitude comparable to that of the most massive loose groups, its faint-end slope is similar to that of loose groups of intermediate stellar mass. Moreover, these authors have shown that the environment of CGs contains more early-type and red galaxies compared to field and loose groups. Finally the X-ray observations of {\it ROSAT}, {\it ASCA}, {\it Chandra} and {\it XMM-Newton} have led to the detection of hot X-ray emitting gas from many CGs \citep{Ponman+96,Mulchaey&Zabludoff99, Fuse&Broming13, Desjardins+13}.

In a compact galaxy system, luminous (hence massive) galaxies are close to one another in projection and are expected to rapidly merge together \citep{Carnevali+81,Schneider&Gunn82,Barnes85,Mamon87}. Several scenarios have been proposed to explain the survival of compact groups against the rapid merging of their galaxies: (i) the appearance of the compact configuration is caused by a chance alignment along the line of sight of galaxies belonging to a parent group \citep{Rose77,Walke&Mamon89} or cosmological filament \citep{Hernquist+95}; (ii) CGs may be transient unbound cores of loose groups \citep{Rose77}; and (iii) CGs of galaxies continually form within a single rich collapsing group, where the dwindling galaxy membership caused by mergers is replenished by new incoming galaxies \citep{Diaferio+94}. Analytical estimates suggest that the replenishment by infall is sufficient \citep{Mamon00_IAU174}. The closest known CG in the Virgo cluster \citep{Mamon89} is almost certainly a product of a chance alignment of galaxies given the redshift-independent distances to its members \citep{Mamon08}. The analysis of semi-analytical models of galaxy formation indicate that roughly two-thirds of CGs selected with HCG criteria are physically dense, while the remaining one-third are caused by chance alignments of galaxies, mostly within virialized groups (\citealp{DiazGimenez&Mamon10}, see also \citealp{Mcconnachie+08}).

If galaxies in CGs are physically close, then galaxies should rapidly merge and form a very luminous, e.g. giant elliptical galaxy \citep{Carnevali+81,Schneider&Gunn82,Mamon86,Mamon87,Barnes89,Dubinski98}. One may then conclude that CGs are the progenitors of the so called fossil groups (FGs, \citealp{Ponman+94}), which are dominated by an isolated giant elliptical galaxy surrounded by X-ray emitting diffuse hot gas \citep{Barnes92,Jones+03}. Unfortunately, it is difficult to observationally distinguish between a real 3D dense environment of CGs and chance alignments within loose groups because of the redshift space distortion uncertainties \citep{Walke&Mamon89}. On the other hand, cosmological $N$-body simulations provide a 3D view of groups and thus allow the study of the nature and properties of CGs \citep{Mcconnachie+08,DiazGimenez&Mamon10} and their evolution in time.

In this paper, we select compact and fossil groups purely on the basis of their halo properties from the Millennium dark matter simulations as well as galaxy properties associated to dark matter halos as characterised based on semi-analytic models (SAMs). We trace back in time both fossil and compact groups, up to $z=1$. Our aim is to (a) investigate the evolution of the CGs in comparison with fossil galaxy groups and (b) address the question of whether there is an evolutionary connection between the two types of galaxy groups, i.e. fossils and compacts, and more specifically if compact groups evolve into FGs.

In Section 2, we present various simulation suites used in this work. In Section 3, we describe the procedure followed to select compact, fossil, and control groups using semi-analytic model catalogs of galaxy formation. We then describe how a mock data has been constructed from a SAM catalog. Out final results are described in Section 4.

%%%%%%%%%%%%%%%%%%%%%%%%%%%%%%%%%%%%%%%%%%%%%%%%%%%%%
%%%%%%%%%%%%%%%%%%%%%%%%%%%%%%%%%%%%%%%%%%%%%%%%%%%%%
\section{Data}
\label{sec2}
Over the last decade, the Cold Dark Matter (CDM) model complemented with the dark energy field $\Lambda$ has been the concordance model for structure formation in the Universe. While the initial growth of density perturbations is linear, the subsequent hierarchical build-up of structures is a highly non-linear process which is only accessible through numerical simulations (e.g., \citealp{Springel+05}). Since the mass component of cold dark matter, which interacts gravitationally, is represented by point particles, the $N$-body simulations can be used to simulate initial perturbations as well as the collapse and formation of structures. 

In this study, we use the Millennium dark matter simulation \citep{Springel+05} along with the publicly available semi-analytic model of \cite{DeLucia&Blaizot07}. While the former provide us with the halo properties of galaxy groups, the latter helps to characterise the physical properties of group's constituent galaxies.

\begin{figure}
\centering
\includegraphics[width=\hsize]{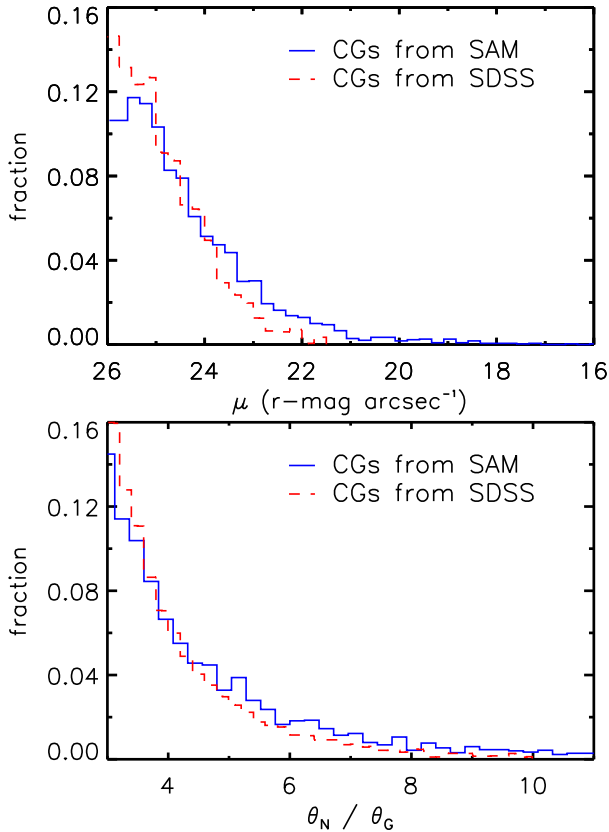}
\caption{Comparison the distributions of observed and simulated compact group parameters: \textit{Upper panel}: $r$-band mean group surface brightness (mag arcsec$^{-2}$). \textit{Lower panel}: ratio of angular distance to nearest neighbor to angular radius of the smallest circumscribed circle. The observed data are those extracted from the SDSS by \citet{Mendel+11} (\textit{red dashed lines}), while the simulated data are extracted from the mock catalog constructed in our study (\textit{blue solid lines}).
}
\label{compare}
\end{figure}

\subsection{The Millennium Simulation} 
The Millennium simulation which is based on the $\Lambda$CDM model, consists of a co-moving periodic box (sides 500$\,h^{-1}$\,Mpc) of 2160$^3$ particles of individual mass $8.6\times10^{8}\, h^{-1}$M$_{\odot}$, and a gravitational softening length of 5$\,h^{-1}$ kpc \citep{Springel+05}. The simulation covers a redshift range from $z=127$ to the present day and is based on an inflationary Universe, leading to a bottom-up hierarchy of structure formation, which involves the collapse and merger of small/dense halos at high redshifts into modern-day observed large virialised systems such as groups and clusters. The cosmological parameters adopted by the Millennium simulation are $\Omega_m = 0.25$, $\Omega_\Lambda = 0.75$, $\Omega_b = 0.045$, $n = 1$ and $\sigma_8 = 0.9$ with the Hubble parameter $h = 0.73$. The locations and velocities of all dark matter particles in the simulation are saved in 64 timesteps roughly logarithmically spread between $z=127$ and $z=0$.

In the Millennium simulation, dark matter halos are identified down to a resolution limit of 20 particles, yielding a minimum halo mass of 1.72$\times 10^{10}\,h^{-1}\,$M$_{\odot}$. Halos in the simulation were found using a Friends-of-Friends (FoF) group finder algorithm, leading to identify halos with over densities of at least 200 times the critical density. Sub-halos are subsequently extracted from the FoF halo with the {\sc SUBFIND} algorithm \citep{Springel+01}. Therefore, for any given halo, a merger tree can be built by output tables at individual epoch. This enables us to hierarchically trace through time the growth of halos and their associated sub-halos within the dark matter simulation.

\subsection{Semi-Analytic Models} 
\label{SAM}
The physical processes in galaxies (e.g. the star formation, thermal evolution of the interstellar medium, the growth of super-massive black holes) occur on very small scales, whereas the evolution of structures happens on cosmological scales. Hence it is very time consuming to follow the details of the physical processes in a single hydrodynamical simulation.

A useful alternative is to rely on semi-analytic models (SAMs) of galaxy formation and evolution \citep{Kauffmann+93,Cole+94}, which evolve galaxies as single entities applying physical recipes to evolve them. SAMs have played an important role in improving our understanding and interpretation of physical processes taking place in galaxies during their evolutions. \cite{DeLucia&Blaizot07} have designed a SAM and run it on the halo merger trees of the Millennium simulation. These merger trees contain sub-halos, and galaxy mergers occur when sub-halos merge into their parent halo or with one another. 

The resulting catalog ({\tt delucia2006a} table in the Millennium database) contains 9 million galaxies at $z = 0$ down to a limiting absolute magnitude of $M_R-5\log h = -15.5$, observed in the $B$, $V$, $R$, $I$ and $K$ filters. There are some differences between the SAM of \citeauthor{DeLucia&Blaizot07} and other SAMs such as \cite{Bower+06}, \cite{Font+08}, and the most recent model of \cite{Henriques+15}. However, for the purpose of our study, which focuses on the evolution of the halos and given the relative success in previous studies \citep[e.g.][]{Gozaliasl+14}, we adopt the SAM developed by \cite{DeLucia&Blaizot07} for our current investigation.

%%%%%%%%%%%%%%%%%%%%%%%%%%%%%%%%%%%%%%%%%%%%%%%%%%%%%
%%%%%%%%%%%%%%%%%%%%%%%%%%%%%%%%%%%%%%%%%%%%%%%%%%%%%
\section{Construction of simulated group samples}
\subsection{Database}
We select all halos, identified by the FoF method, from the dark matter halo catalog of the GAVO database\footnote{http://gavo.mpa-garching.mpg.de/Millennium/} at $z = 0$ with halo mass $M(R_{200})\geq 10^{13}\,h^{-1}\,\rm M_{\odot}$. Group galaxy members associated to these halos were selected from the \cite{DeLucia&Blaizot07} SAM output ({\tt delucia2006a} table) of GAVO. The halo mass threshold was applied to ensure that the progenitors of the present day galaxy groups are indeed groups at $z\sim1.0$ with at least four galaxy members \citep{Dariush+10}. The evolution of each group was then followed, from $z = 0$ to $z =1.0$, in 23 discrete snapshots equally spaced in $\log z$, by matching \texttt{haloID}s to their progenitors at earlier epochs. In addition, the BGG position of each galaxy group as well as the position of its host dark matter halo were used to identify group members by using \texttt{fofID} and \texttt{haloID} keys in the SAM catalog. This helps to retrieve optical properties of group member galaxies from the semi-analytic galaxy catalog at any redshift.

\subsection{Mock redshift-space catalog}
A mock redshift-space catalog can be constructed from the real space SAM using the algorithm described in \citet{Blaizot+05}, but without the box transformations (translation, rotation and flipping) or replication. To do so we have performed the following steps:
\begin{enumerate}
    \item We place the observer at one of the vertices of the simulation box and use the line-of-sight to the observer as the time line of the light cone.
    \item We convert the Cartesian coordinate system to celestial coordinates (i.e. RA and Dec).
    \item We compute the redshift of each galaxy following \cite{Duarte&Mamon15}. To perform this, we solve $d_{\rm comov}(z_{\rm cos}) = \sqrt{X^2+Y^2+Z^2}$ for the cosmological redshift $z_{\rm cos}$. And then obtain the galaxy redshift by applying the Doppler shift at the cosmological redshift by $1+z = \sqrt{(1+\beta)/(1-\beta)}\,(1+z_{\rm cos})$, where $\beta$ is the line-of-sight component of the peculiar velocity divided by the velocity of light: $\beta = \bm{v_{\rm p}}/c\cdot \bm{d}/d$.
    \item The luminosity distance ${D_{L}}$ and therefore the apparent magnitude of each galaxy are then computed. 
    \item Finally, k-corrections are applied to correct the apparent magnitudes, using the color-based method of \cite{Chilingarian+10}.
\end{enumerate}

\subsection{Samples of simulated groups}
\label{sec3}
We select samples of mock fossil and compact groups from the Millennium simulation. In addition, following the previous studies of FGs (e.g. \citealp{Dariush+07,Dariush+10}), we also select a mock control sample whose properties are similar to normal galaxy groups \citep{Smith+10}, without being fossil or compact. We choose to work with redshift-space samples, since CGs can only be selected in redshift-space.

\subsubsection{Fossil groups}
Using the conventions introduced by \citet{Jones+03}, FGs are systems having the following properties: 
\begin{description}
    \item [X-rays] A spatially extended halo of X-ray emission with a minimum bolometric luminosity of $L_{X,{\rm bol}} \geq 10^{42}\, h^{-2}$\,erg~s$^{-1}$. 
    \item [dominant galaxy] the difference in magnitudes between the two brightest group galaxies within $0.5\,R_{200}$ should have $\Delta m_{12} = r_2-r_1 \geq 2$. Here, $R_{200}$ is the radius of the sphere centered on the halo in which the critical density is 200 times the mean density of Universe.
\end{description} 

According to the scaling relation between halo mass and X-ray luminosity, an X-ray luminosity of $L_{X,{\rm bol}} \geq 10^{42}\,h^{-2}$\,erg~s$^{-1}$ in the Millennium simulation corresponds to the halo mass of $M_{200}\geq 10^{13}h^{-1}\,\rm M_{\odot}$ \citep{Dariush+07,Dariush+10}. Hence, halos of our selected FGs meet a minimum mass limit (see \S~3.1). In \S~\ref{foss}, the other samples of FGs will be described in more detail.

We compile two samples of FGs. In the first one, FGs are selected based on the DeLucia2006a \citep{DeLucia&Blaizot07} real-space catalog in the Millennium 3D volume and in the second one, fossils are identified in the mock catalog (redshift-space). 

Some real-space FGs may no longer be classified as fossils in redshift-space due to the projection effects. In other words, the FG criterion ($\Delta m_{12} \ge 2$ within $0.5\,R_{200}$) is a more conservative constraint in redshift space than in real space. To explore this, we trace back the halo mass evolution of what we call {\it hybrid fossil groups}, i.e. the real-space FGs that fail to be identified as redshift-space FGs, against those identified as fossils in the mock catalog (i.e. redshift sample).

\subsubsection{Compact groups}
\label{compactSelect}
CGs are selected from the mock catalog, using an automated search algorithm similar to the one described by \citet{Hickson82}. In doing so, the following criteria were applied to the FoF galaxy groups:
\begin{description} 
    \item [Population] $N \geq 4$ galaxies within 3 magnitudes from the brightest ($m_{\rm b}$) in the $R$ band 
    \item [Compactness] mean surface brightness $\mu_R \leqslant 26$ mag arcsec$^{-2}$, within smallest circumscribed circle of angular diameter $\theta_{\rm G}$ containing the galaxy centers; 
    \item [Isolation] distance to nearest neighboring galaxy in same magnitude range building a larger circle of angular diameter $\theta_{\rm N} \geqslant 3\,\theta_{\rm G}$. In other words, the concentric annulus of angular radii between $\theta_{\rm G}$ and $3\,\theta_{\rm G}$ must be devoid of galaxies in the [$m_{\rm b}$,$m_{\rm b}$+3] magnitude range, and is called the \emph{isolation annulus}.
\end{description}

\begin{figure}
\centering
 \includegraphics[width=\hsize]{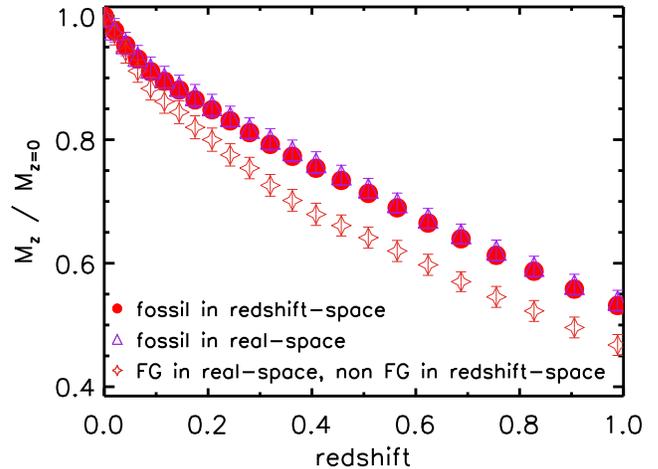}
\caption{Geometric mean mass assembly histories of the most massive progenitors of fossil groups. Comparison of the real-space fossil groups (\textit{violet open triangles}), the redshift-space fossil groups (\textit{red filled circles}) and the hybrid fossil groups (real-space fossils not present in redshift-space fossil group sample, \textit{dark red stars}). The errors on the mean MAH are computed according to equation~(\ref{errMAH}).
}
\label{fosmock}
\end{figure} 

\begin{figure}
\includegraphics[width=\hsize]{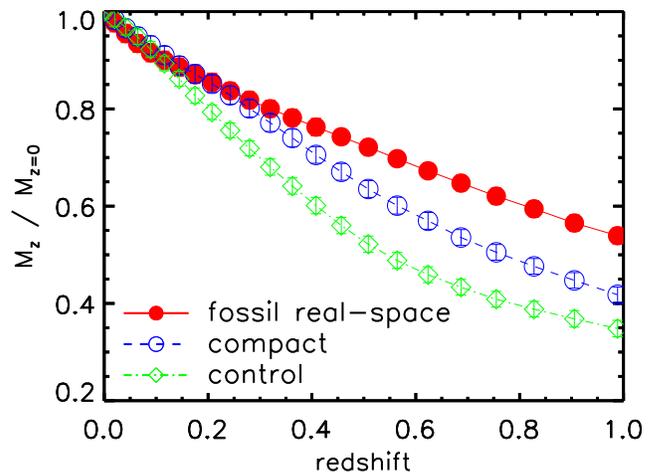}
\caption{Geometric mean mass assembly histories of the most massive progenitors of real-space fossil groups (\textit{red filled circles}), compact groups (\textit{blue open circles}), and our control groups (\textit{green open diamonds}) The mass evolution of compact groups is between those of fossil groups and control groups. The errors on the mean MAH are computed according to equation~(\ref{errMAH}).}
\label{mtrace}
\end{figure}

In Fig.~\ref{compare}, the surface brightness $\mu$ and the angular diameter of the ratio of the largest to the smallest concentric circles $\theta_{\rm N}/\theta_{\rm G}$ are compared to those estimated by \cite{Mendel+11}, from a compilation of CGs, based on the SDSS DR6 data. As is clear in Fig.~\ref{compare}, there is a fair agreement between the observed and simulated distributions of the group surface brightnesses on one hand and of distance to closest neighbor in units of group sizes on the other hand, although the simulated CGs are more likely to be very compact ($\mu<24\,\rm mag\,arcsec^{-2}$) and very isolated ($\theta_{\rm N}/\theta_{\rm G} > 5$) than the ones that extracted by \citeauthor{Mendel+11}

\subsubsection{Control sample}
A sample of control groups have also been selected as a representative of predominantly young galaxy groups. Control groups are systems with $\Delta m_{12} \leq 0.5$ within $0.5\, R_{200}$. In addition, they do not belong to either fossil or compact groups.

\begin{table}[h]
\tabcolsep 3pt
    \begin{center}
    \caption{Mock group samples
    \label{t1a}}
    \begin{tabular}{lr}
        \toprule
        Group sample & Number\\ 
        \hline \bottomrule
        All (non-compact, real space) & 51538 \\ 
        Fossil (real space sample) & 10150 \\ 
        Fossil (redshift space sample) & 8751 \\ 
        Fossil (hybrid: in real space but not in redshift space) & 1854 \\
        Control  & 10625 \\
        Compact  & 2330 \\ 
        \hline 
        \bottomrule
    \end{tabular}
    \end{center}
    \noindent {\sc Notes}: All groups are extracted from the $z$=0 \citep{DeLucia&Blaizot07} semi-analytic model output and have halo mass $M(R_{200}) \geq 10^{13}\, h^{-1}\,\rm M_{\odot}$.
\end{table} 

Table~\ref{t1a}, summarises the number of groups of different types in the Millennium mock catalog (and in the Millennium 3D catalog in case of real space sample). The number of real-space FGs is 15 per~cent higher than the corresponding number of redshift-space FGs. Also, 82 per~cent of real-space FGs (8296) are also in the redshift-space FG sample, which means that 18 per~cent of real-space FGs (1854) do not meet the fossil criteria in redshift space.

\begin{figure}
\includegraphics[width=\hsize]{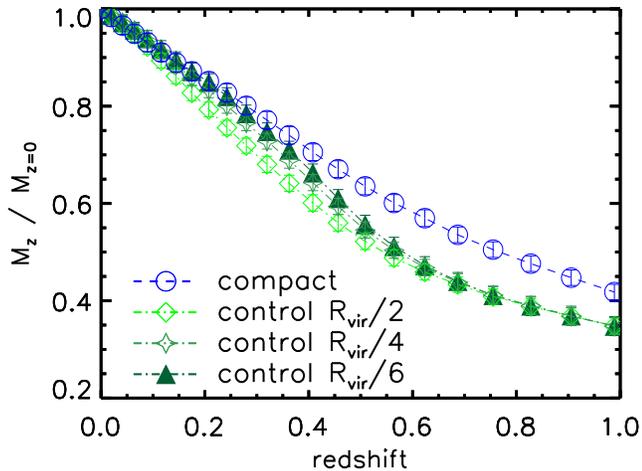}
\caption{Geometric mean mass assembly histories of the most massive progenitors of compact groups (\textit{blue open circles}) compared with several control samples: the full control sample (\textit{green open diamonds}), control groups within $R_{\rm vir}/4$ (\textit{mid green stars}) and within $R_{\rm vir}/6$ (\textit{dark green filled triangles}). Decreasing the maximum extent of control groups in units of their virial radius leads to increasingly similar (slower) evolution as compact groups.
}
\label{control2}
\end{figure}

\begin{figure} 
\centering
\includegraphics[width=\hsize]{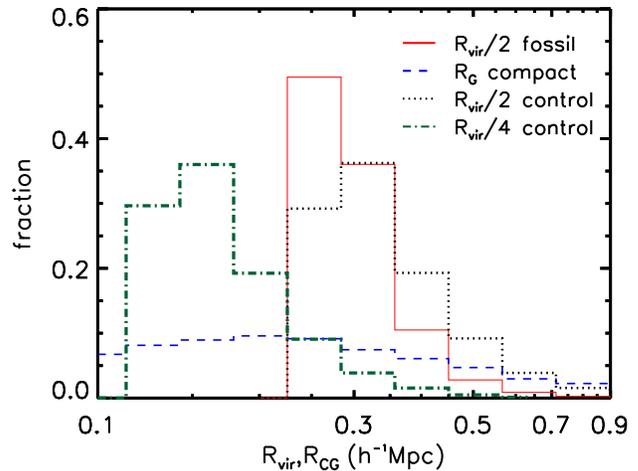}
\caption{Distribution of compact group physical radii (\textit{blue dashed histogram}), half virial radii of fossil groups (\textit{solid red histogram}), and fractional virial radii of control groups (\textit{green histograms}) for $R_{\rm vir}/2$ (\textit{dotted}) and $R_{\rm vir}/4$ (\textit{dash-dotted}) with evenly logarithmic bins of $(\log 2)/3$ scale.
}
\label{rvir}
\end{figure}

%%%%%%%%%%%%%%%%%%%%%%%%%%%%%%%%%%%%%%%%%%%%%%%%%%%%%
%%%%%%%%%%%%%%%%%%%%%%%%%%%%%%%%%%%%%%%%%%%%%%%%%%%%%	
\section{Results}
\label{sec4}
\subsection{Mass assembly history of fossil groups}
\label{foss}

\begin{figure*}
\includegraphics[width=\hsize]{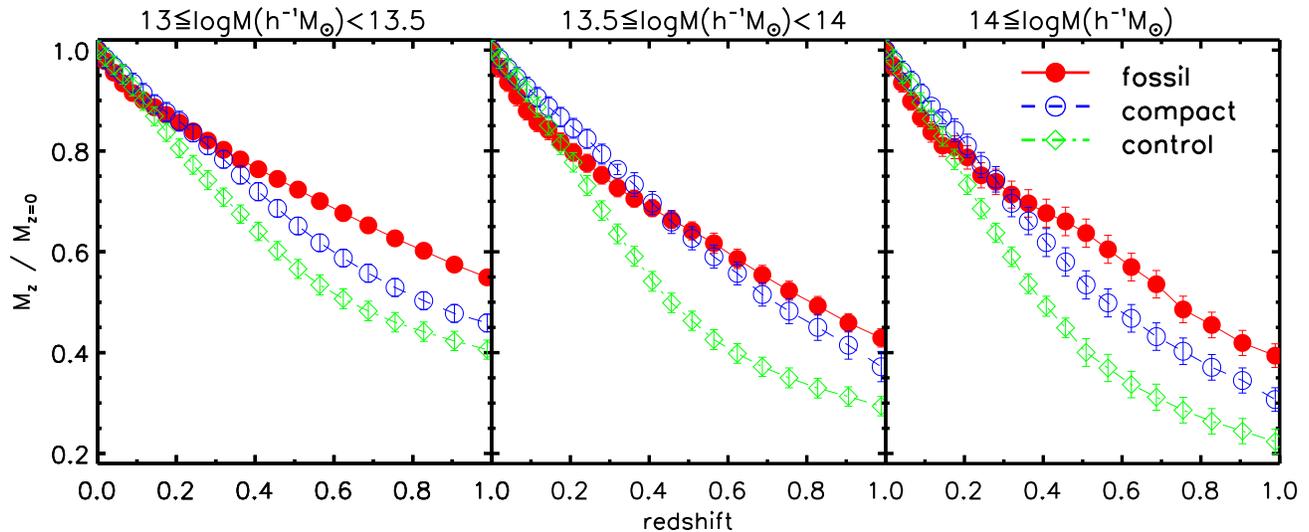}
\caption{Geometric mean mass assembly histories of the most massive progenitors of compact (\textit{blue open circles}), fossil (\textit{red filled circles}) and control groups (\textit{green open diamonds}) in three different bins of final halo mass, increasing from left to right.
}
\label{mass2}
\end{figure*}

In previous studies of the evolution of FGs from galaxy formation simulations, FGs had been selected in real space \citep{Dariush+07,Dariush+10}. But since our aim is to investigate the connection between CGs and FGs by studying their \emph{mass assembly histories} (MAHs) in the Millennium Simulation, and since CGs are selected in redshift space, we first identify fossils in redshift-space and then verify whether their geometric mean MAH is consistent with earlier results based on the 3D approach.

We measure the errors $\epsilon$ on the mean MAH, $\left \langle M(z)/M_{(z=0)}\right \rangle$, by considering the square root of the sum of the squared statistical errors, with the cosmic variance. The statistical error on the mean is $\sigma_{\rm stat} = \sigma/N^{1/2}$, where $\sigma$ is the standard deviation of the MAH for the groups of the considered type at the considered redshift and $N$ is the number of these groups at that redshift. We measure the cosmic variance by dividing the simulation box into 8 cubic sub-boxes of half the box size, and writing that the cosmic variance is 1/8th of the variance $\sigma_{\rm 8-sub-boxes}^2$ of the means of the 8 boxes. The errors $\epsilon$ on the mean MAH then satisfy

\begin{equation}
\epsilon^2 = \sigma_{\rm stat}^2 + \sigma_{\rm CV}^2 = {\sigma^2\over N} + {\sigma_{\rm 8-sub-boxes}\over 8}\ .
\label{errMAH}
\end{equation}

At low masses, the cosmic variance is of the same order as the statistical uncertainties, but at high masses, statistical uncertainties are dominant. The same error estimation was done for all group samples.

Fig.~\ref{fosmock} compares the mean mass assembly histories (MAHs), $M_z/M_{z=0}$ of the different FG samples. One clearly sees that while real-space and redshift-space FGs have the same mean MAH, suggesting a high degree of similarity between the mean MAHs in both samples.

\subsection{Mass assembly history of compact groups}
The MAH of each group was then traced from $z = 0$ to $z = 1$ in 23 discrete snapshots, equally spaced in $\log z$. We then consider the \emph{mean MAH}, $\left\langle M_z/M_{z=0}\right\rangle$, averaging over all $z$=0 groups for all CG, FG and control group samples.

Fig.~\ref{mtrace} compares the mean MAHs of fossil, compact and control groups from $z=0$ to $z=1.0$. From this figure, it is clear that, in agreement with previous studies \citep[e.g.][]{Dariush+07}, FGs have assembled a larger fraction of their final halo mass at earlier epochs in comparison to control groups. The mean MAH in CGs falls in between those of fossil and control groups. At low redshifts $z \lesssim 0.2$, the mean MAH of Compact groups is very similar to that of FGs and of control groups, with a surprisingly slightly faster recent growth of FGs relative to CGs and control groups for $z \lesssim 0.12$, i.e. 1.5 Gyr. 

To better understand the observed discrepancy in the mean MAH of compact and control groups, given the differences in their selection criteria, we select two more samples of control groups by applying the same magnitude gap criterion, but instead of considering the first- and second-ranked galaxies to be within $R_{\rm vir}/2$, we select them to be within $R_{\rm vir}/4$ and $R_{\rm vir}/6$ and trace back their most massive progenitors. As Fig.~\ref{control2} shows, decreasing the radius within which the first two brightest group galaxies are selected from $R_{\rm vir}/2$ to $R_{\rm vir}/6$ (logically the next step after $R_{\rm vir}/4$ should be $R_{\rm vir}/8$, but there are too few group members at this radius, hence we test $R_{\rm vir}/6$ instead) causes the mean MAH in control groups to approach the one seen in CGs at $z \lesssim 0.5$. This suggests that, in comparison to control groups, the observed trend in CGs is dictated by its compact configuration. However, the distribution of the angular sizes of the CGs (angular radii, $\theta_{\rm G}$, of the smallest circumscribed circle), shown in Fig.~\ref{rvir}, covers a wide range of values including those estimated within $R_{\rm vir}/2$ in FGs, as well as the ones derived for control groups within $R_{\rm vir}/2$, $R_{\rm vir}/4$ Hence, CGs seem to behave heterogeneously compared to the fossil and control samples.

To explore whether the results depends upon the $z$=0 halo mass, we have repeated the above procedure in three different bins of final halo mass. Fig.~\ref{mass2} shows that, while the mean MAHs of fossil, compact and control groups each depend on final halo mass (except, surprisingly, the FGs between the intermediate and high mass bins), the hierarchy of mean MAHs between fossil, compact and control groups remains the same as in Figs.~\ref{mtrace} and \ref{control2}: at $0.3 < z < 1$, the most massive progenitors of the $z$=0 control groups grow faster than those of the CGs, which in turn grow faster than those of the FGs; but at $z<0.1$, the mean growths of the most massive progenitors are very similar between the three classes of groups, with the most massive progenitors of $z$=0 FGs showing a more slightly more rapid growth than the corresponding progenitors of compact and control groups.

\subsection{Concentration}
\begin{figure}
	\centering
	\includegraphics[width=\hsize]{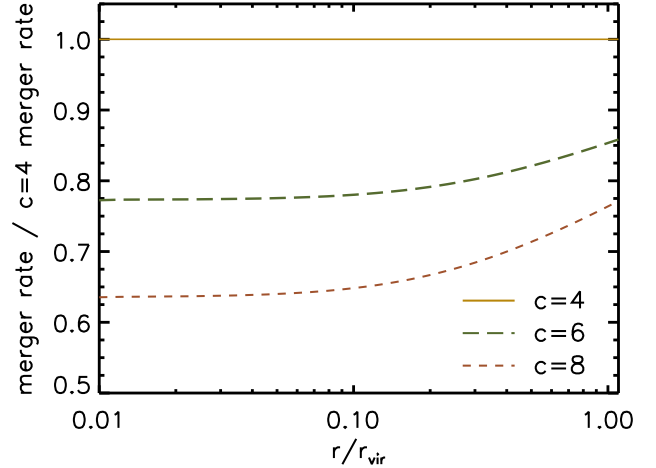}
	\caption{Ratio of merger rate $(\propto \rho/v_{\rm c}^3)$ to that of $c$=4 halo for NFW models, assuming circular orbits with the \cite{Chandrasekhar43} formula.}
	\label{mergerrates}
\end{figure}

Large magnitude gaps in galaxy groups are generally believed to be caused by galaxy mergers, as the most massive (luminous) galaxy grows by mergers, usually by merging with the 2nd-ranked galaxy \citep{Mamon87}. The (group) halo concentration can change the rate at which galaxy mergers occur in a galaxy group: according to the \cite{Chandrasekhar43} formula, the rate of mergers by dynamical friction roughly scales as $\rho/v_{\rm circ}^3$, which for \citep{Navarro+96} (NFW) models at fixed virial quantities leads to a slightly \emph{lower} rate of mergers for higher concentrations at given ratio of radius over virial radius, as shown in Fig.~\ref{mergerrates}. In high concentration NFW halos, the higher densities are offset by even higher 3rd powers of the circular velocities. 

It is well known that, for given final halo mass, higher concentration halos assembled earlier \citep{Wechsler+02}. Therefore, if FGs assemble earlier than CGs, we expect that their $z$=0 concentrations should be higher than those of CGs. 

We describe the dark matter with an NFW density profile:
\begin{equation}
    \rho_{\rm NFW} = \frac{\rho_{s}}{(r/r{_s})(1 + r/r{_s})^2} \ ,
    \label{nfw}
\end{equation}
where $r_{\rm s}$ is the scale radius (where the logarithmic slope of the density profile is equal to --2), while $\rho_{s} = 4 \rho(r_{\rm s})$ is a characteristic density. The `standard' concentration of the halo can be defined as $c \equiv c_{200} = r_{200}/r_{s}$, where $r_{200}$ is the enclosing mean density of 200 times the critical density. At $z=0$, relaxed halos in $\Lambda$CDM cosmological simulations have median concentrations decreasing, with slope $\simeq -0.1$, from 6.5 to 5 for $\log (h M_{200}/\rm M_\odot)$ increasing from 13 to 14.5 (e.g., \citealp{Navarro+97,Neto+07,Prada+12}). Unrelaxed halos have concentrations roughly one-third lower \citep{Neto+07}. Finally, the distribution of concentrations of relaxed halos of given halo mass is roughly lognormal with a dispersion of 0.1 dex \citep{Neto+07}.

Unfortunately, the scale radius $r_{\rm s}$ is not readily available in the Millennium database. Instead, the concentration could be defined on the ratio of the virial radius to the radius containing half the mass enclosed within the virial sphere: $c_{\rm h} = r_{200}/r_{\rm h}$. But this definition of concentration does not capture well the standard concentration, since for NFW models, $c_{\rm h} \simeq 1.45\,c^{0.28}$ for reasonable values of $c$, i.e., $c_{\rm h}$ ranges in the small interval from 2.0 to 2.6 for $c$ varying from 3 to 8. 

We therefore followed \cite{Prada+12} in defining the concentration of galaxy groups from 
\begin{equation}
    {v_{\rm max}\over v_{200}} =0.465 \sqrt{c\over \ln(c+1)-c/ (c+1)} \ ,
    \label{vmaxoverv200}
\end{equation}

where $v_{\rm max}$ is taken from the Millennium simulation. The ratio $v_{\rm max}/v_{200}$ is a U-shaped function of $c$ that reaches a minimum of unity at $c = 2.163$, which corresponds to the radius where the circular velocity curve is maximum. Equation~(\ref{vmaxoverv200}) can thus only be solved if $v_{\rm max}/v_{200} > 1$, which ensures that the solution of equation~(\ref{vmaxoverv200}) for $c$ has two roots. We thus solve equation~(\ref{vmaxoverv200}), adopting the greater of the two solutions for $c$, i.e. the one with $c>2.163$. Since, the maximum circular velocity is reached at a radius of $2.163\,r_{\rm s}$ and since our virial radii satisfy $r_{\rm vir} = c\,r_{\rm s}\geq 2.163\,r_{\rm s}$, we are guaranteed that the radius of maximum circular velocity is smaller than the virial radius. Finally, we compute the virial velocity from $v_{200} = \sqrt{GM_{200} / r_{200}}$ where $G = 43.01\,\rm (km\,s^{-1})^2\, Mpc\, (10^{10}\,\rm M_{\odot})^{-1}$ is Newton's gravitational constant.

\begin{figure}
	\centering
	\includegraphics[width=\hsize]{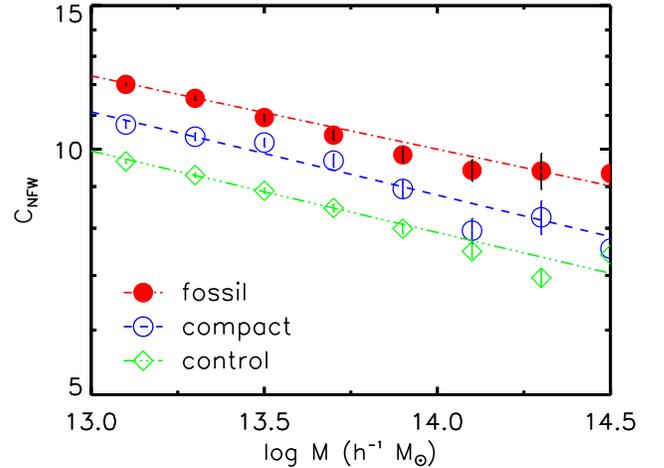}
	\caption{Halo concentration parameters $c = r_{200}/r_{\rm s}$ (derived by solving eq.~[\ref{vmaxoverv200}] for $c$) as a function of halo mass, for fossil, compact and control groups at $z=0$. The symbols represent the median concentration in bins of the halo mass, and the error bars are the standard deviation among halos of the same mass divided by the square root of the halo number in the bin masses.}
	\label{concen}
\end{figure}

In Fig.~\ref{concen}, we compare the concentration-mass relations of fossil, compact and control groups. At given halo mass, the concentration parameters of CG halos tends to be 10 per~cent smaller than those of FGs, but 10 per~cent larger than those of control groups. In all three classes of galaxy groups, the correlation between the mass and concentration is well defined. A power-law provides a good description of the median concentration as a function of halo mass, and we find:
\begin{eqnarray}
    c_{200}^{\rm FG} &=& 12.16\left(\frac{h M_{200}}{10^{13}M_{\odot}}
    \right)^{-0.08}\,
\label{cFG}\\
    c_{200}^{\rm CG} &=& 11.14\left(\frac{h M_{200}}{10^{13}M_{\odot}}
    \right)^{-0.1} \,
\label{cCG}\\
    c_{200}^{\rm control} &=& 9.95\left(\frac{h M_{200}}{10^{13}M_{\odot}}
    \right)^{-0.1} \ .
\label{ccontrol}
\end{eqnarray}

The differences in the concentration-mass relation normalizations among the three classes of groups is clearly statistically significant, given the small uncertainties on the means in Figure~\ref{concen}. We tested this by noting a slope of $-0.1$ is an acceptable power law index for all these classes, and that with this slope, the normalizations at $\log M/{\rm M}_\odot = 14$ are 0.83$\pm$0.02, 0.79$\pm$0.02, and 0.78$\pm$0.01 for FGs, CGs, and control groups, respectively. This clearly shows the statistically significant differences.

\cite{Wechsler+02} showed that the dispersion in the mass concentration is fairly large, therefore the question arises how can CGs lead to systematically lower concentrations than FGs? To address this question, as showed in Fig.~\ref{mtrace}, the assembly time of CGs are later than the FGs. Therefore, it is not surprising that the concentrations of the CGs are systematically lower than those of the FGs. In addition, as the middle panel of Fig.~\ref{mass2} shows, the assembly time of CGs are similar to FGs in the mass bin $13.5 \leq \log\left(hM/{\rm M}_{\odot} \right) < 14$ of the final halo mass, and so the concentrations of CGs are closer to those of FGs at $\log\left(hM/{\rm M}_{\odot} \right) < 13.5$ and 13.7 in Fig.~\ref{concen}, which confirm the above picture.

In some X-ray or lensing surveys, massive clusters have unusually high concentration. For instance, using X-ray data, \cite{Buote+07} found that $\log (h\,M/{\rm M}_{\odot}) = 13.5$ groups have \emph{dark matter} concentration $c = 15$ defined using the virial radius, i.e., $c \simeq 11$ when the outer radius is taken to be $r_{200}\simeq r_{\rm vir}/1.35$ for NFW models of reasonable concentration.\footnote{Fitting a model to the total mass density profile instead of the dark matter one leads \cite{Buote+07} to obtain much higher concentrations, which is caused by the stellar component dominating the inner regions \citep{Mamon&Lokas05}.} 

Using kinematical modelling, \cite{Mamon07_Chile} found that, dynamically hot ($\sigma_{v}>$300 kms$^{-1}$) X-ray selected groups have $c>10$, while cold ($\sigma_{v}\leq$300 kms$^{-1}$) groups mostly are located at $c<5$, where the concentrations are measured for the total distribution of mass. Our high standard concentrations are thus consistent with the dynamically hot X-ray selected groups.

There is an issue whether the existence of a gap in the galaxy luminosity function in `fossil' systems is the `last state' or is a `transitory phase' in the group evolution? To address this issue many works have been done. For instance, \cite{2008MNRAS.386.2345V} showed that many groups will go through a `fossil phase' which typically will be ended by new infalling satellites from the environment and cause leaving the phase. Furthermore, \cite{2011MNRAS.416.2997C}, found no significant difference in the central galaxy properties in the FGs and non-FGs. These findings was consistent with the analysis carried out by \cite{Dariush+10}, who showed that regardless of the redshift at which FGs are identified, after $\sim$4 Gyr, more than $\sim$90 per cent of them become non-FGs. Moreover, beyond the extent of 7.7 Gyr (time interval between $z$ = 0-1) very few groups retain a 2 mag gap between their two brightest galaxies. This provides clear evidence that the FGs are simple groups that temporarily are in a `fossil phase'. On the other hand, in current work, we find that FGs have a higher concentration than the control groups which may be due to the fact that in regular systems the concentration is higher than the merging systems, and FGs are known to have avoided recent mergers.

%%%%%%%%%%%%%%%%%%%%%%%%%%%%%%%%%%%%%%%%%%%%%%%%%%%%%
%%%%%%%%%%%%%%%%%%%%%%%%%%%%%%%%%%%%%%%%%%%%%%%%%%%%%
\subsection{Evolution of parameters} 
\begin{figure}
\includegraphics[width=\hsize]{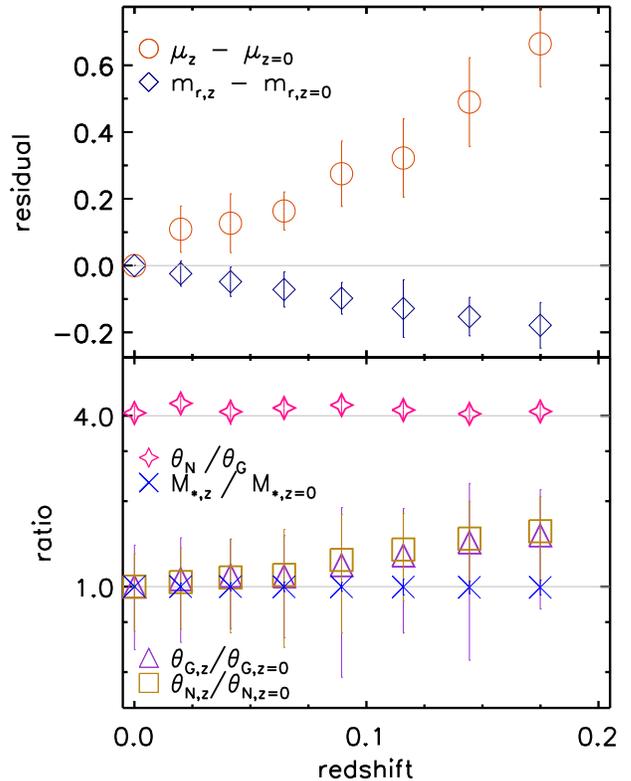}
\caption{Evolution of the properties of $z$=0 compact groups: surface brightness $\mu_{z} - \mu_{z=0}$ (\textit{dark orange circles}), apparent magnitude $m_{r,z} - m_{r,z=0}$ (\textit{dark blue diamonds}), angular outer radius of isolation annulus $\theta_{N,z}/\theta_{N,z=0}$ (\textit{violet triangles}), angular radius of group $\theta_{G,z}/\theta_{G,z=0}$ (\textit{golden squares}), ratio of outer to inner radii of isolation annulus $\theta_{N}/\theta_{G}$ (\textit{pink stars}) and stellar mass evolution of galaxies within group radii (\textit{blue asterisks}).
}
\label{tracef}
\end{figure}

We now analyze the time evolution of CG parameters i.e. group mean surface brightness $\mu_R$, group isolation ring outer ($\theta_{\rm N}$) and inner ($\theta_{\rm G}$) angular radii (see \S~\ref{compactSelect}), by tracing back the parameters of the present day ($z=0$) CGs to the epoch when they are no longer compact. Since most CGs disappear within the first few snapshots before $z=0$, the sample size becomes too small for reasonable statistical measurements. For example, among 2330 CGs at $z=0$, only 6 groups remain compact at $z=0.3$. We therefore trace back the CG parameters only to $z=0.2$ when $\sim$140 compact groups remain. 

Fig.~\ref{tracef} shows that from $z=0.2$ to $z=0$, CGs evolve with slightly increasing surface brightness (a decrease of 0.1 in surface magnitude). This slightly higher surface brightness at $z=0$ could be explained by today's CGs being either more luminous or smaller (or both) than their CG progenitors at $z=0.2$. As shown in Fig.~\ref{tracef}, the projected radius of the CG ($\theta_{\rm G}$) is $\sim1.5$ times larger at $z-0.2$ than its radius at the current epoch $(z=0)$. If the group luminosity were fixed, this decrease in CG angular radius would lead to the surface magnitude decreasing by $-0.9$, hence the group luminosity must simultaneously fade by $\approx 0.8$ magnitude. Also, the angular isolation radius of the group ($\theta_{\rm N}$) experiences a similar decrement over time. Therefore, the ratio of $\theta_{\rm N}/\theta_{\rm G}$ remains almost unchanged (\textit{open stars}).

%In Fig.~\ref{tracef} the $\Delta\mu$ is jumped from 0.1 to 0 between $z=0.02$ to $z=0$ while it is much more constant in time at higher $z$. To address this rising, in the CG's tracing back from 2330 CGs at $z=0$, only $\sim$500 groups are compact in the higher snapshot ($z=0.02$). This shows that about 80\% of today CGs have been assembled in two last snapshots which mostly caused by satisfying the isolation criteria by a wet merging (gas-rich) of the farthest member of the compact core to their neighbor bright galaxies. Considering constant $\theta_{G}$ for last two snapshots (the lower panel of Fig.~\ref{tracef}) and 0.1 magnitudes rise in surface brightness, the total group magnitude should be increased to see the jump in $\Delta\mu$ evolution. The upper panel of Fig.~\ref{tracef} clearly shows an increase in the group total magnitude of the CGs. While the lower panel of Fig.~\ref{tracef} shows a constant evolution of stellar mass ($-2.5\log\left( M_{\star}/L \right)$ where $M_{\star}$ is total stellar mass of group). Therefore starbursts in a wet merger at last step of CG evolution is enhanced the total magnitude of group core while their stellar mass stayed constant.

According to Fig.~\ref{tracef}, the evolution of CGs since z=0.175 occurs at nearly constant group stellar mass (it increases by only 0.6 per cent). This slow increase of stellar mass is consistent with the negligible growth of the total galaxy stellar mass (summed over all progenitors) found by \cite{DeLucia&Blaizot07} for brightest cluster galaxies (their Fig. 7). During the same period, the CG total luminosities dim by 20 per cent, roughly as expected from passive evolution of a constant mass stellar population. This luminosity dimming is too small to compete with the decrease of CG size by a factor 1.5, which is the main contributor to the increase of the mean surface brightness by a factor 1.9.

\begin{figure}
\includegraphics[width=\hsize]{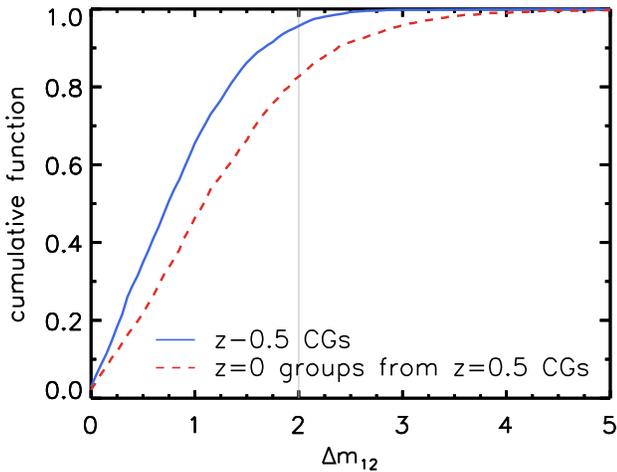}
\caption{Cumulative distribution function of magnitude gaps of the compact groups selected at $z=0.5$ (\textit{blue solid line}) and the compact groups traced forward to $z=0$ (\textit{red dashed line}).
}
\label{deltamup}
\end{figure}

Furthermore, we select CGs at $z=0.5$ (selected in the same way as described in Sec.~\ref{sec4}), trace their evolution forward to $z=0$, and follow the evolution of the magnitude gap (selected within the half virial radius). According to Fig.~\ref{deltamup}, the fraction of groups with large ($\Delta m_{12}>2$) magnitude gaps increases with time from 5 per~cent to 20 per~cent, and thus some CGs turn into fossils by $z=0$. But this process does not significantly contribute to the population of the present-day FGs, as the fossil phase does not survive for a long time and a galaxy group may go through the fossil phase several times during its evolution \cite[e.g.,][]{Dariush+10}.

%%%%%%%%%%%%%%%%%%%%%%%%%%%%%%%%%%%%%%%%%%%%%%%%%%%%%
%%%%%%%%%%%%%%%%%%%%%%%%%%%%%%%%%%%%%%%%%%%%%%%%%%%%%
\subsection{Connection between compact and fossil groups}
\begin{table}
    \begin{center}
    \caption{Link between compact and fossil groups}
    \hrule
    \hrule
    \begin{tabular}{lcc}
        Conversion & Observer & Fraction \\
        \toprule
        Simultaneous FG & box center & 0.03 $\pm 0.02$ \\
        Progenitor of FG was CG & box center & 0.23 $\pm 0.03$ \\
        Progenitor of FG was CG & any & 0.36 $\pm 0.02$ \\
        CG turns into FG & box center & 0.41 $\pm 0.03$ \\
    \end{tabular}
    \hrule
    \label{t2}
    \end{center}

    Notes:
    	Row~1: probability of a galaxy group to be simultaneously fossil (FG) and compact (CG) at $0\leq z \leq 1$;
    	Row~2: probability that a progenitor of a $z$=0 fossil group is a compact group identified at $0<z \leq 1$;
    	Row~3: probability that the brightest group galaxy of a $z$=0 fossil group is the most luminous galaxy of a compact group identified at $0<z \leq 1$, for any observer in the simulation box;
    	Row~4: probability that a compact group at $z=0.5$ turns into a fossil group by $z=0$.
\end{table}

\begin{figure}
\includegraphics[width=\hsize]{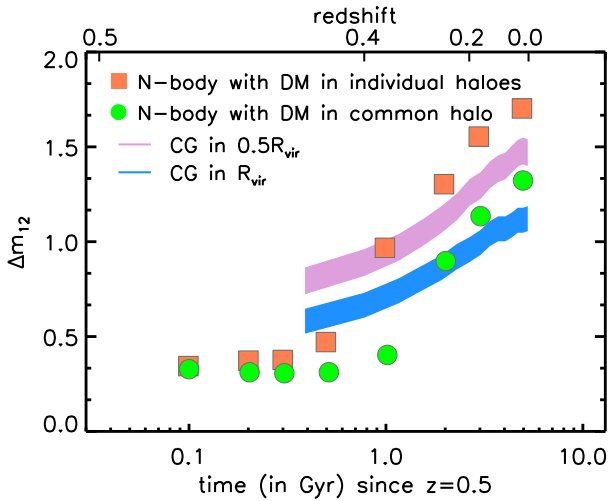}
\caption{Evolution of the magnitude gap for various sets of compact groups: compact groups selected at $z=0.5$ and traced forward to $z=0$, where the gap is measured for the 2 most luminous galaxies within the virial radius (\emph{blue strip}) or within half the virial radius (\emph{plum strip}). Also shown are compact groups of 8 galaxies run in the $N$-body simulations of virialized groups by \cite{Mamon87}, with either individual dark matter halos (\emph{orange squares}), or a common dark matter halo (\emph{green circles}).
}
\label{dm12t}
\end{figure}

It is intriguing to know if any connection exists between compact and fossil groups. For instance, do all CGs evolve into FGs? In other words, are FGs the end products of galaxy mergers in CGs \citep{Mamon87,Barnes92,Jones+03}. Also, do all FGs go through a compact phase before evolving into the form it has at the present epoch? 

To address these questions, we study the progenitors of the present-day FGs to determine, statistically, what fraction of them went through a compact phase at earlier epochs. More precisely, FGs are initially identified in redshift space in the mock catalog at $z=0$, and at every previous snapshot, we check whether any of their progenitors are compact (they may no longer be a FG) by applying the CG selection criteria at each redshift slice. We iterate this analysis up to $z=1$, unless a progenitor of the $z$=0 FG is found to meet the CG criteria at a given $z<1$ snapshot. We find that only $\approx$23 per~cent of the present-day FGs were also CGs at some stage in the past (see Table~\ref{t2}). Note that the probability that a galaxy group satisfies both compact and fossil criteria simultaneously is only $\lesssim$3 per~cent in any snapshot within the redshift range $0\lesssim z \lesssim 1$. We also trace forward the CGs selected at $z=0.5$ up to $z=0$ and check whether any of these groups turn into FGs. We also proceed until a CG meets the FG criteria at a given $0.5<z<0$ snapshot. We find that $\sim$ 41 per~cent of CGs become FGs during their evolution. Therefore, most CGs do not have enough time for their magnitude gap to grow above 2 magnitudes. This can also be deduced from Fig.~\ref{dm12t}, which indicates that it takes over 10 Gyr since $z=0.5$ for the mean gap of CGs to grow above 2 magnitudes. 

The snapshots are spaced by 350 Myr at $z=1$ and 260 Myr at $z=0$. Could the CG phase be shorter than the time resolution of the Millennium simulation? Using several tens of $N$-body simulations, \cite{Mamon87} concluded that dense groups of galaxies lose their CG appearance in projection in typically 750 Myr if the dark matter is around the individual galaxies, and nearly 4 times longer if the dark matter is in a common envelope. Therefore the CG phase lasts longer than the time between the Millennium snapshots, hence we should have missed very few CGs.

Note that, in the above exercise, the selection of groups in the redshift space as described in Sec.~\ref{sec4}, is based on an observer located at the center of the simulation box. Therefore the BGG of a FG which is selected based on the `absolute' magnitude, is not necessarily the brightest galaxy of a CG (which selected based on the `apparent' magnitude) while applying the compact criteria. Thus, we can also move around the position of an observer in the simulation box such that a progenitor of the BGG of a FG is always the brightest galaxy in the CG found among the progenitors of the FG (rather than considering a fixed observer's position). In this case, we find that around $\approx$36 per~cent of $z$=0 FGs were also CGs at epochs $0<z\leq 1$.

If only a minority of FGs have gone through the CG phase, and if the large magnitude gaps in FGs are signs of more rapid mergers than in control groups, one must conclude that such rapid merging can occur in groups that fail to meet the Hickson CG criteria. Since the galaxy merger rates are higher in dense systems (where the dynamical times are shorter), one expects that the HCG compactness criterion is not the issue, but rather the HCG isolation criterion. In other words, the progenitors of FGs should have sufficiently dense cores for rapid merging to occur and thus for the magnitude gap to grow, but these cores are not necessarily isolated from their surroundings. To address this idea, we trace back the progenitors of FGs that have CG population of at least 4 galaxies and go through a compact phase but not considering the isolation criteria up to $z=1$. We then find that $\sim$72 per~cent of today FGs satisfy the HCG compactness and population, but fail the isolation criteria.

Fig.~\ref{dm12t} shows the evolution of the magnitude gap, $\Delta m_{12}$, for the 3123 CGs identified in the Millennium simulation at $z=0.5$ and in sets of 50 $N$-body simulations of virialized dense groups of 8 `halos' by \cite{Mamon87}. In these simulations, each galaxy was represented by a single particle, with an additional particle for the intragroup background. Each particle had structure, mass and energy, which were exchanged between particles during mergers, rapid collisions, and lost to the background (particle) through dynamical friction (which puffed up the background particle). As expected, $\Delta m_{12}$ increases rapidly in time, and \cite{Mamon87} concluded that this rise in magnitude gap is the consequence of galaxy mergers. The linear trend of the semi-log plot of Fig.~\ref{dm12t} indicates an exponential increase of the magnitude gap in CGs. Fig.~\ref{dm12t} also shows that the magnitude gap grows a little faster after 2~Gyr when we select the two most luminous galaxies within half the virial radius (instead of within $r_{\rm vir}$). The slower growth of the gaps of CGs selected at $z=0.5$ in the (cosmological) Millennium simulation relative to that in the idealized simulations of virialized groups of \cite{Mamon87} probably arises from luminous infalling galaxies that fill the gap.

%%%%%%%%%%%%%%%%%%%%%%%%%%%%%%%%%%%%%%%%%%%%%%%%%%%%%
%%%%%%%%%%%%%%%%%%%%%%%%%%%%%%%%%%%%%%%%%%%%%%%%%%%%%
\section{CONCLUSIONS}
\label{sec5}
In this work, we extracted fossil and compact groups from the outputs of the \cite{DeLucia&Blaizot07} semi-analytical model, run on the dark matter cosmological Millennium Simulation. This allowed us to analyze the mass assembly history of compact and fossil groups and explore the connection between the two classes of groups. Our major conclusions from the analyses can be summarized as follows:

\begin{itemize}
    \item As many as $\sim$18 per~cent of fossils in the Millennium 3D
      catalog do not meet the fossil criteria in the mock catalog because of
      projection effects.

    \item Fossils are older than compact groups, since by $z = 1$, fossils
      have assembled more than $\sim$55 per~cent of their $z$=0 halo mass,
      compared to only $\sim$40 per~cent for $z=1$ compact groups
      (Fig.~\ref{mtrace}).

    \item The mass accretion history of compact groups in the mass range
      $13<\rm log(M/\rm M_{\odot})<13.5$ is very similar to that of control
      groups, but in the halo mass range of $13.5<\rm log(M/\rm M_{\odot})
      <14$, it is more similar the corresponding evolution of fossil
      groups. But in general it seems that CGs follow the FG evolution more
      closely than they follow that of control groups
      (Fig.~\ref{mass2}).

    \item Compact groups and fossils both show trends of halo concentration
      slightly decreasing with halo mass, but, at given halo mass,
      the concentrations of compact groups are roughly 10\% lower than those of
      fossil groups (Fig.~\ref{concen}).
      
    \item From $z=0.2$ to $z=0$ the angular radii of the inner and outer
      circles of the isolation annuli around the groups are compressed by a
      factor of $1.5$ with time. This ``compression'' of compact groups comes
      with a dimming of their luminosity, with their surface brightness only
      slightly dimming in time (Fig.~\ref{tracef}).
      
    \item Finally, while as many as 3/4 of fossil groups have appeared
      compact since $z=1$, most of these compact systems fail the compact
      group isolation criterion, hence are not truly 
      compact groups as defined here, leaving only $\sim 23\% - 36\%$ of fossil
      meeting the compact group criteria between $z=1$ and $z=0$.
      Therefore, compact and fossil groups are not intimately related classes of
      groups of galaxies. 

    \item The magnitude gap in compact groups selected at $z=1$ increases
      exponentially in time (Fig.~\ref{dm12t}), but takes $\sim 10$ Gyr to
      grow to 2 magnitudes on average. This explains why only a minority of CGs (41 percent) turn into FGs at $z=0$.

\end{itemize}

It therefore seems that compact groups constitute a specific class of groups, rather than being part of the general evolutionary path of groups that may lead to the formation of fossils. Our future work will focus on the observational properties of compact and fossil groups using current galaxy surveys. Our aim will be to understand any possible link between compact and fossil groups.

%%%%%%%%%%%%%%%%%%%%%%%%%%%%%%%%%%%%%%%%%%%%%%%%%%%%%
%%%%%%%%%%%%%%%%%%%%%%%%%%%%%%%%%%%%%%%%%%%%%%%%%%%%%
\acknowledgments{We thank Dr. Eugenia D\'iaz-Gim\'enez for her guidance and useful discussions. The Millennium simulation used in this paper was carried out by the Virgo Supercomputing Consortium at the Computing Centre of the Max-Planck Society in Garching. The semi-analytic galaxy catalog \citep{DeLucia&Blaizot07} is publicly available at http://gavo.mpa-garching.mpg.de, and we thank Gabriela De Lucia and Jeremy Blaizot for allowing public access for the outputs of their very impressive semi-analytical models of galaxy formation. This research made use of the ``K-corrections calculator'' service \citep{Chilingarian+10} available at http://kcor.sai.msu.ru .}

%%%%%%%%%%%%%%%%%%%%%%%%%%%%%%%%%%%%%%%%%%%%%%%%%%%%%
%%%%%%%%%%%%%%%%%%%%%%%%%%%%%%%%%%%%%%%%%%%%%%%%%%%%%
\bibliographystyle{yahapj}
\bibliography{AminFarhang}

\end{document}